\documentclass[9pt]{article}

\usepackage{url}
\usepackage{caption}
\usepackage[usenames,dvipsnames]{xcolor}
\usepackage{ragged2e}
\usepackage[cmex10]{amsmath}
\usepackage{float}
\usepackage{graphicx}
\usepackage{amssymb,amsfonts}
\usepackage{fullpage}
\usepackage{times}
\usepackage{overpic}

\graphicspath{{figures/}}
\DeclareGraphicsExtensions{.pdf,.png,.jpg}

\newcommand{\reffig}[1]{Figure\,\ref{#1}}
\newcommand{\refeq}[1]{equation\,(\ref{#1})}

\makeatletter
\renewcommand\@biblabel[1]{#1 }

\def\@maketitle{%
  \newpage\spacing{1}\setlength{\parskip}{10pt}%
    \begin{center}%
	{\large\bfseries\noindent\sloppy \textsf{\@title} \par}%
    {\noindent\sloppy \@author}%
	\end{center}%
}

\renewcommand{\section}{\@startsection {section}{1}{0pt}%
    {-6pt}{1pt}%
    {\reset@font \large \bfseries}%
    }
\renewcommand{\subsection}{\@startsection {subsection}{2}{0pt}%
    {-6pt}{1pt}%
    {\reset@font \normalsize \bfseries}%
    }
\makeatother

\newcommand{\spacing}[1]{\renewcommand{\baselinestretch}{#1}\large\normalsize}

\AtBeginDocument{}

\begin{document}

\title{Optimizing Hybrid Spreading in Metapopulations}

\author{
Changwang Zhang$^{1,2,4,*}$, Shi Zhou$^{1,*}$, Joel C. Miller$^{5,6,7}$, Ingemar J. Cox$^1$, and Benjamin M. Chain$^3$\\
\vspace{2ex}
$^1$Department of Computer Science, $^2$Security Science Doctoral Research Training Centre, and $^3$Division of Infection and Immunity, University College London, UK. $^4$ School of Computer Science, National University of Defense Technology, Changsha, China. $^5$ School of Mathematical Sciences, $^6$School of Biological Sciences, and $^7$Monash Academy for Cross \& Interdisciplinary Mathematics, Monash University, Melbourne, Victoria, Australia.
$^*$Author for correspondence  (changwang.zhang.10@ucl.ac.uk or s.zhou@ucl.ac.uk)
}

\maketitle

\begin{abstract}
Epidemic spreading phenomena are ubiquitous in nature and society. Examples include the spreading of diseases, information, and computer viruses. Epidemics can spread by \textit{local spreading}, where infected nodes can only infect a limited set of direct target nodes and \textit{global spreading}, where an infected node can infect every other node. In reality, many epidemics spread using a hybrid mixture of both types of spreading. In this study we develop a theoretical framework for studying hybrid epidemics, and examine the optimum balance between spreading mechanisms in terms of achieving the maximum outbreak size. We show the existence of \textit{critically} hybrid epidemics where neither spreading mechanism alone can cause a noticeable spread but a combination of the two spreading mechanisms would produce an enormous outbreak. Our results provide new strategies for maximising beneficial epidemics and estimating the worst outcome of damaging hybrid epidemics.

\end{abstract}


\section*{Introduction}

Epidemic spreading phenomena are ubiquitous in nature and society. Examples include the spreading of infectious diseases within a population, the spreading of computer viruses on the Internet, and the propagation of information in society. Understanding and modelling the dynamics of such events can have significant practical impact on health care, technology and the economy. Various spreading mechanisms have been studied \cite{Newman_Book_2010,Keeling_Eames_2005}. The two most common mechanisms are {\it local spreading}, where infected nodes only infect a limited subset of  target nodes \cite{Pastor-Satorras_Vespignani_2001}; and {\it global spreading}, where nodes are fully-mixed such that an infected node can infect any other node \cite{Anderson_1991,Newman_Book_2010}.  In reality, many epidemics use {\it hybrid spreading}, which involves a combination of two or more spreading mechanisms. For example the computer worms Conficker \cite{Shin_2012} and Code-Red \cite{Moore_Shannon_claffy_2002} can send probing packets to targeted computers in the local network or to any randomly chosen computers on the Internet.

Early relevant studies investigated epidemics spreading in populations whose nodes mix at both local and global levels (``two levels of mixing'') \cite{Ball_1997}. These early studies \cite{Ball_1997} did not incorporate the structure of the local spreading network, assuming both local and global spreading are fully-mixed. Since the introduction of network based epidemic analysis \cite{Pastor-Satorras_Vespignani_2001,Newman_Book_2010}, hybrid epidemics have been studied in structured populations \cite{Vazquez_2007}, in structured households \cite{Ball_2012, House_Keeling_2008, Ma_2013}, and by considering networked epidemic spreading with ``two levels of mixing'' \cite{Ball_Neal_2008, Kiss_Green_Kao_2006, Estrada_2011}. A number of studies \cite{Watts_2005,Colizza_Vespignani_2007,Mata_2013,Min_2013, Keeling_2010, Apolloni_2014} have also considered epidemics in metapopulations, which consist of a number of weakly connected subpopulations. The studies of epidemics in clustered networks \cite{Miller_2009,Tildesley_2010,Volz_2011} are also relevant.
Much prior work on hybrid epidemics has focused on the impact of a network's structure on spreading.

Most previous studies were about what we call the \textit{non-critically} hybrid epidemics where a combination of multiple mechanisms is not a necessary condition for an epidemic outbreak. In this case, using a fixed total spreading effort, a hybrid epidemic will always be less infectious than an epidemic using only the more infectious one of the two spreading mechanisms \cite{Kiss_Green_Kao_2006, Wang_Jin_2013}. However, many real examples of hybrid epidemics suggest the existence of \textit{critically} hybrid epidemics where a mixture of spreading mechanisms may be more infectious than using only one mechanism.

In this paper we investigate whether, and if so when, hybrid epidemics spread more widely than single-mechanism epidemics. We propose a mathematical framework for studying hybrid epidemics and focus on exploring the optimum balance between local and global spreading in order to maximize outbreak size. We demonstrate that hybrid epidemics can cause larger outbreaks in a metapopulation than a single spreading mechanism. 

Our results suggest that it is possible to combine two spreading mechanisms, each with a limited potential to cause an epidemic, to produce a highly effective spreading process. Furthermore, we can identify an optimal tradeoff between local and global mechanisms that enables a hybrid epidemic to cause the largest outbreak. Manipulating the balance between local and global spreading may provide a way to improve strategies for disseminating information, but also a way to estimate the largest outbreak of a hybrid epidemic which can pose serious threats to Internet security. 

\section*{The Hybrid Epidemic (HE) Model}

\begin{figure}\centering\small
\includegraphics[width=0.8\textwidth]{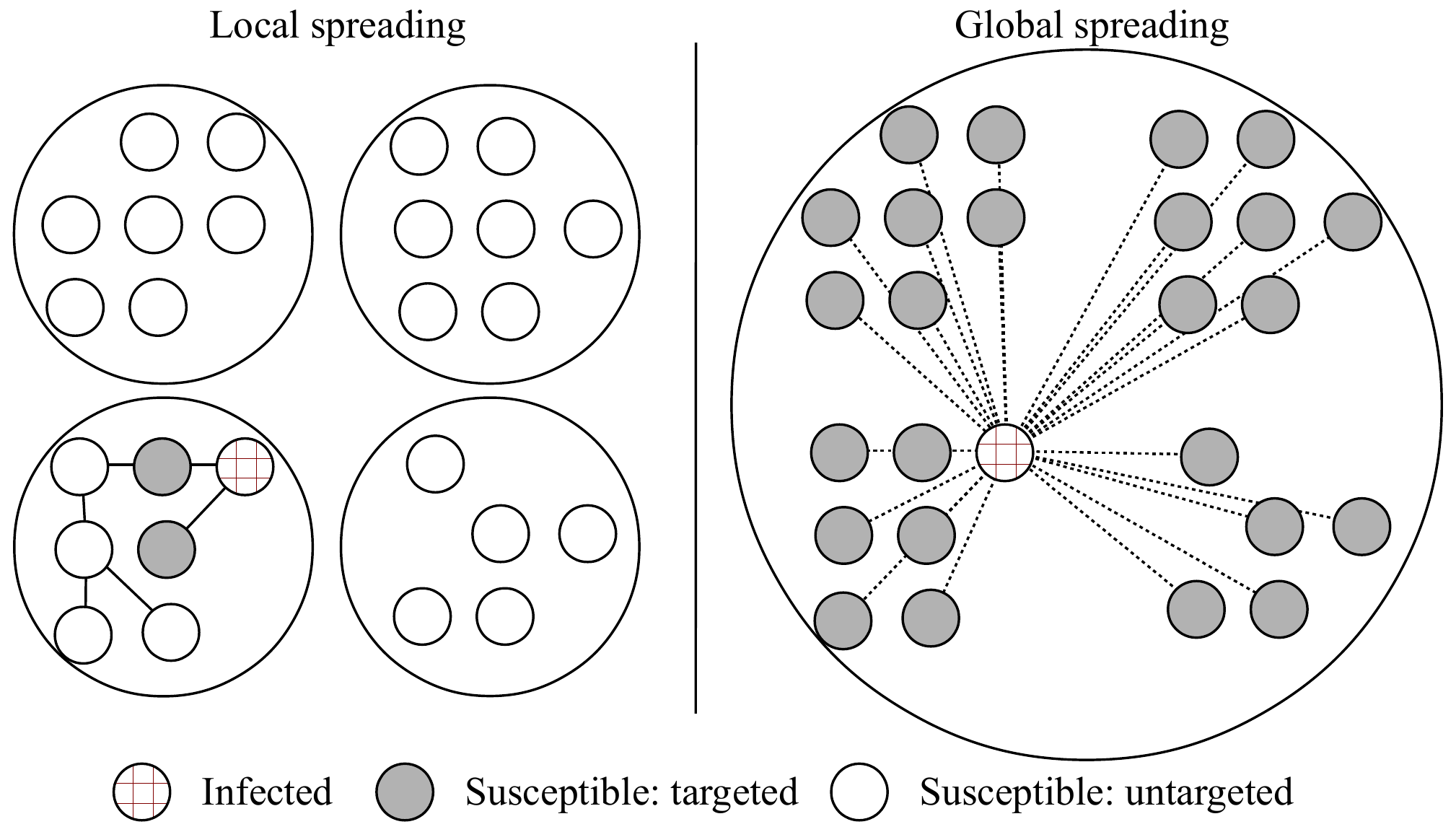}
\caption{\label{fig-l-g}Hybrid epidemic spreading in a metapopulation. At each time step, an infected node has a fixed total spreading effort which must be allocated between local spreading and global spreading. The proportion of spreading effort spent in local spreading is  $\alpha$ and that in global spreading is $1-\alpha$. Local spreading occurs between infected and susceptible nodes that are connected in individual subpopulations; global spreading happens between an infected node and any susceptible node in the metapopulation.}
\end{figure}

Here we introduce a model for hybrid epidemics  in a {\it metapopulation}, which consists of a number of subpopulations. Each subpopulation is a collection of densely or strongly connected nodes, whereas nodes from different subpopulations are weakly connected. As illustrated in \reffig{fig-l-g}, our model considers two spreading mechanisms: 1) local spreading where an infected node can infect nodes in its subpopulation and 2) global spreading, where an infected node can infect all  nodes in the metapopulation. In our model each subpopulation for local spreading can be either fully-mixed or a network. For mathematical convenience, we describe each subpopulation as a network and represent a fully-mixed subpopulation as a fully connected network. Note that our definition of metapopulation is different from the classical metapopulation defined in ecology where subpopulations are connected via flows of agents \cite{Colizza_Vespignani_2007, Keeling_2010}.

Our model considers hybrid epidemics in which at each time step, an infected node has a fixed total spreading effort which must be allocated between the two spreading mechanisms. Let the hybrid tradeoff, $\alpha$, represent the proportion of spreading effort spent in local spreading. The proportion of global spreading effort is $1-\alpha$. A tunable $\alpha$ enables us to investigate the interaction and the joint impact of the two spreading mechanisms on epidemic dynamics, ranging from a completely local spreading scenario (with $\alpha=1$) to a completely global spreading scenario (with $\alpha=0$). For example computer worms like Conficker \cite{Shin_2012} and Code-Red \cite{Moore_Shannon_claffy_2002} can conduct both local and global probes but the average total number of probes in a time unit is fixed.

We consider the hybrid epidemic spreading in terms of the Susceptible-Infected-Recovered (SIR) model \cite{Newman_Book_2010, House_2012}, where each node is in one of three states: \textit{susceptible} (s), \textit{infected} (i), and \textit{recovered} (r). At each time step, each infected node spreads both locally and globally; it infects 1) each directly connected nodes in the same subpopulation with rate $b_1=\alpha\beta_1$ and 2) each susceptible node in the metapopulation with rate $b_2=(1-\alpha)\beta_2$. $\beta_1$ is the  local infection rate when all spreading effort is local ($\alpha=1$). And $\beta_2$ is the  global infection rate when all spreading effort is global ($\alpha=0$). Each infected node recovers at a rate $\gamma$, and then remains permanently in the recovered state. A node can infect other nodes and then recover in the same time step.

\section*{Hybrid Spreading In A Single-Population \label{sec:hone}}

Before we analyse hybrid spreading in a metapopulation, we study a relatively simple case where the epidemic process takes place in a {\it single population}. That is, there is only one population, where local spreading is via direct connections on a network structure and global spreading can reach any node in the population.

Here we extend the system in \cite{Miller_2012} for the analysis. The system in \cite{Miller_2012} was proposed to analyse single-mechanism based epidemics for the continuous time case. Here we extend the system to analyse 1) hybrid epidemics, and 2) for the discrete time case. We calculate the probability that a random test node $u$ is in each state: susceptible $s(t)$, infected $i(t)$, and recovered $r(t)$.

We denote $p(k)$ as the probability that a node has degree (i.e. number of neighbours) $k$. The generating function~\cite{Newman_Strogatz_Watts_2001} of degree distribution $p(k)$ is defined as $g_0(x)=\sum_{k} p(k) x^k$. Let $p_n(k)$ represent the probability that a random neighbour of $u$ has $k$ neighbours. We assume the network is {\it uncorrelated}: the degrees of the two end nodes of each link are not correlated (i.e. independent from each other) \cite{Newman_Book_2010}. In an uncorrelated network $p_n(k)=p(k)k/\langle k \rangle$ \cite{Newman_Book_2010}.

Let $\theta(t)$ be the probability that a random neighbour $v$ has not infected $u$ through local spreading. Let $\vartheta(t)$ be the probability that a random node $w$ has not infected $u$ through global spreading. Suppose $u$ has $k$ neighbours, the probability that it is susceptible is $s_k(t)=\vartheta(t)^{n-1}\theta(t)^k$ where $n$ is the total number of nodes in the population. Then by averaging $s_k(t)$ over all degrees, we have,
\begin{equation}\label{eq-st}
s(t)=\vartheta(t)^{n-1} \sum_k p(k) \theta(t)^k=\vartheta(t)^{n-1}g_0(\theta)
\end{equation}
The probability $\theta$ can be broken into three parts: $v$ is susceptible at $t$, $\phi_s $; $v$ is infected at $t$ but has not infected $u$ through local spreading, $\phi_i$; $v$ is recovered at $t$ and has not infected $u$ through local spreading, $\phi_r$. Neighbour $v$ can not be infected by $u$ and itself, then $\label{eq-phis}
\phi_s=\vartheta^{n-2} \sum_k p_n(k) \theta^{k-1}=\vartheta^{n-2} g_0'(\theta)/g_0'(1)
$. In a time step, neighbour $v$ 1) infects $u$ with rate $b_1\phi_i$ through local spreading and 2) recovers without infecting $u$ through local spreading at rate $\gamma(1-b_1)\phi_i$, i.e. after every time step: $(1-\theta)$ increases by $b_1\phi_i$ and $\phi_r$ increases by $\gamma(1-b_1)\phi_i$. The increase rate of $\phi_r$ here, $\gamma(1-b_1)\phi_i$, is different from that ($r\phi_i$) in the original system in \cite{Miller_2012}. Because the original system was designed for the continuous time case, and in the discrete time case in this paper, neighbour $v$ can infect $u$ and recovers at the same time step. Given that $\phi_r$ and $1-\theta$ are both approximately 0 in the beginning ($t=0$), we have $\phi_r=\gamma(1-b_1)(1-\theta)/b_1$. Then
\begin{equation}\label{eq-phii}
\phi_i=\theta-\phi_s-\phi_r = \theta - \vartheta^{n-2} \frac{g_0'(\theta)}{g_0'(1)} - \frac{\gamma(1-b_1)}{b_1} (1-\theta)
\end{equation}
For global spreading, the probability $\vartheta$ can also be broken into three parts: $w$ is susceptible at $t$, $\varphi_s$; $w$ is infected at $t$ but has not infected $u$ through global spreading, $\varphi_i$; $w$ is recovered at $t$ but has not infected $u$ through global spreading, $\varphi_r$. Using a similar derivation process, we have $\varphi_s=\vartheta^{n-2} \sum_k p(k) \theta^k=\vartheta^{n-2} g_0(\theta)
$ and $\varphi_r=(1-\vartheta)\gamma(1-b_2)/b_2$, and
\begin{equation}\label{eq-varphii}
\varphi_i=\vartheta-\varphi_s-\varphi_r=\vartheta-\vartheta^{n-2} g_0(\theta) - \frac{\gamma(1-b_2)}{b_2} (1-\vartheta)
\end{equation}
When the epidemic stops spreading, $\phi_i=0$ and $\varphi_i=0$. By setting $\phi_i=0$ in \refeq{eq-phii} we get
\begin{equation}\label{eq-vartheta-n-2}
\vartheta^{n-2} = \frac{g_0'(1)}{g_0'(\theta)}(\theta+\frac{\gamma(1-b_1)}{b_1}\theta-\frac{\gamma(1-b_1)}{b_1})
\end{equation}
Substituting \refeq{eq-vartheta-n-2} and $\varphi_i=0$ into \refeq{eq-varphii}, we have
\begin{equation}\label{eq-vartheta}
\vartheta=w(\theta)=\frac{g_0'(1)(\theta+\theta\gamma(1-b_1)/b_1-\gamma(1-b_1)/b_1)g_0(\theta)/g_0'(\theta)}{1+\gamma(1-b_2)/b_2}+\frac{\gamma(1-b_2)/b_2}{1+\gamma(1-b_2)/b_2}
\end{equation}
By setting $\phi_i=0$ and substituting \refeq{eq-vartheta} in \refeq{eq-phii} we have
\begin{equation}\label{eq-theta}
\theta = \frac{\vartheta^{n-2} g_0'(\theta)/g_0'(1)+\gamma(1-b_1)/b_1}{1+\gamma(1-b_1)/b_1} = f(\theta)
\end{equation}
Then $\theta_{\infty}$ - stationary value of $\theta$ is a fixed point of $f(\theta)$. 

\subsection*{Threshold Condition}

$f(\theta)$ has a known fixed point of $\theta=1$ which represents no epidemic outbreak. We test the stability of this fixed point. By substituting \refeq{eq-phii} and \refeq{eq-vartheta} into $d\theta/dt=-b_1\phi_i$, setting $\theta=1+\epsilon$ and take the leading order (Taylor Series), we have $d\epsilon/dt=\epsilon h$ and
%
%
%
\begin{equation}\label{eq-h}h=\dfrac{b_1 A+b_2 B+b_1 b_2 C-\gamma^2}{b_2+\gamma-b_2\gamma}\end{equation}
where $A=\gamma^2+g_0''(1) \gamma/g_0'(1)-\gamma$, $B=\gamma^2+n \gamma-3 \gamma$, and $C=-\gamma^2-g_0''(1) \gamma/g_0'(1)-n \gamma+4 \gamma-n g_0''(1)/g_0'(1)+3 g_0''(1)/g_0'(1)+n g_0'(1)-2 g_0'(1)+n-3$.Then $\epsilon=Ce^{ht}$ where $C$ is a constant. When $h$ is negative, $|\epsilon|$ gradually decreases and approaches 0 as $t$ increases; while when $h$ is positive, $|\epsilon|$ gradually increases and approaches $+\infty$ with the increase of $t$. That is the fixed point $\theta=1$ turns from stable to unstable when $h$ changes from negative to positive. A more rigorous analysis would need to consider the fact that when a small amount of disease is introduced, the fixed point at $\theta=1$ is moved slightly. However, the stability analysis we do here is sufficient to determine whether epidemics are possible for arbitrarily small initial infections. Further details are in \cite{Miller_2014}.  The threshold condition for an epidemic outbreak is then $h>0$:
\begin{equation}\label{eq:threshold}
h(\beta_1, \beta_2, \gamma, \alpha, p(k))>0,
\end{equation}
%
%
This epidemic threshold  represents an condition which, when \textit{not} satisfied, results in an epidemic that vanishes exponentially fast~\cite{Anderson_1991, Castellano_Pastor-Satorras_2010}. There are two special cases. 
\begin{itemize}
\item
For completely local spreading ($\alpha=1,b_1=\beta_1, b_2=0$), the threshold reduces to $\beta_1g_0''(1)/[g_0'(1)(\beta_1+\gamma-\gamma\beta_1)]>1$. Here $g_0'(1)=\langle k\rangle$ and $g_0''(1)=\langle k^2 \rangle-\langle k \rangle$ where $\langle k\rangle$ is the average degree of the network and $\langle k^2 \rangle$ is the average degree square of the network \cite{Newman_Book_2010}. In the methods section, we show that this threshold agrees with previous threshold results \cite{Newman_2002} for single-mechanism epidemics spreading on networks for the discrete time case. For infinite scale-free networks, we have $(\langle k^2 \rangle-\langle k \rangle)/\langle k \rangle \rightarrow \infty$ such that the threshold `vanishes' (i.e. $\infty>1$ is always satisfied), in agreement with previous observation~\cite{Pastor-Satorras_Vespignani_2001,Newman_Book_2010}.

\item
For completely global spreading ($\alpha=0,b_1=0,b_2=\beta_2$), the threshold reduces to $\beta_2(n-3+\gamma)/\gamma>1$, and when $n$ is large it is approximate to $\beta_2n/\gamma>1$. $\beta_2n/\gamma$ is the {basic reproduction number}, $R_0$, for single-mechanism epidemics spreading in a fully mixed population~\cite{Anderson_1991}. $R_0$ is the average number of nodes that an infected node can infect before it recovers. Thus the threshold is equivalent to $R_0>1$, in agreement with previous work \cite{Anderson_1991}. 

\end{itemize}

\subsection*{Final Outbreak Size}

The final outbreak size, $r_{\infty}$, is the fraction of nodes that are recovered when all epidemic activities cease, i.e.~when all nodes are either recovered or susceptible. When $t\rightarrow\infty$, the probability that a node is infected $i(t) \rightarrow 0$. Thus $r_{\infty}=1-s_{\infty} =1-\vartheta_{\infty}^{n-1}g_0(\theta_{\infty})$ and 
\begin{equation} \label{eq-rinf0}
r_{\infty}=1-w(\theta_{\infty})^{n-1}g_0(\theta_{\infty})
\end{equation}
where the value of $\theta_{\infty}$ can be numerically calculated by conducting the fixed-point iteration of \refeq{eq-theta}. Equation \ref{eq-rinf0} can be viewed as a function of the hybrid epidemic parameters and the network degree distribution. To be noted here, for completely global spreading ($\alpha=0,b_1=0,b_2=\beta_2$),  $\theta_{\infty}$ can not be calculated from \refeq{eq-theta} (because $b_1=0$). In this case, $\theta_{\infty}=1$, $g_0(\theta_{\infty})=1$, and $r_{\infty}=1-\vartheta_{\infty}^{n-1}$ where $\vartheta_{\infty}$ can be obtained by setting $\varphi_i=0$, $g_0(\theta)=1$ and solving the \refeq{eq-varphii} in the rage $0<\vartheta<1$.

\subsection*{Evaluation}

Numerical simulations were performed to verify the above theoretical predictions for hybrid epidemics in a single population.  We consider three  topologies for local spreading in the single-population: (1) a fully connected network which represents a fully mixed population; (2) a random network with Poisson degree distribution, which is generated by the Erd\H{o}s-R\'enyi (ER) model \cite{ER_1959} with average degree 5; and (3) a scale-free network with a power-law degree distribution $p_k \sim 2m^2k^{-3}$, which is generated by the configuration model \cite{Newman_Book_2010} with the minimum degree $m=3$. Each of these  networks has 1000 nodes. At the beginning, 5 randomly selected nodes are infected and all others are susceptible.

\begin{figure}
\centering\small
\includegraphics[width=0.8\textwidth]{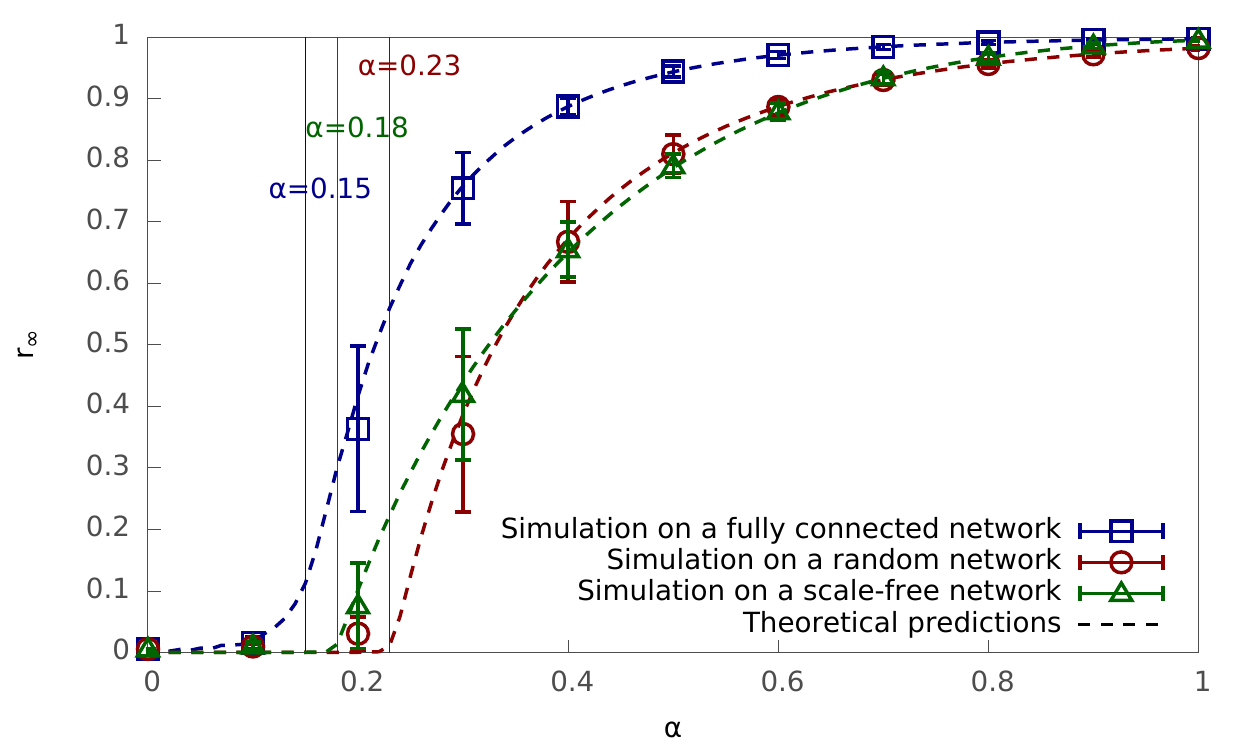}
\caption{\label{fig-threshold}Theoretical predictions and simulation results for hybrid epidemics in a single-population. The final outbreak size $r_{\infty}$ is shown as a function of the hybrid tradeoff $\alpha$. Three network topologies are considered: (1) a fully connected network (i.e. fully mixed); (2) a random network with an average degree of 5; (3) a scale-free network with a power-law degree distribution $p_k \sim 2m^2k^{-3}$ which is generated by the configuration model \cite{Newman_Book_2010} with the minimum degree $m=3$. The population has 1000 nodes. The  global infection rate $\beta_2=10^{-4}$ and recovery rate $\gamma=1$ are the same for epidemics on these three types of networks. The  local infection rate $\beta_1$ is $6\times10^{-3}$ for epidemics on the fully connected network; and it is $0.8$ for epidemics on the random  and scale-free networks. Initially 5 random nodes are infected. Simulation results are shown as points and theoretical predictions of \refeq{eq-rinf0} are dashed curves. The simulation results are averaged over 1000 runs with bars showing the standard deviation. The epidemic threshold values of $\alpha$ are predicted by \refeq{eq:threshold} and marked as vertical lines.}
\end{figure}

We run simulations for different values of $\alpha\in[0, 1]$. We set the global infection rate $\beta_2=10^{-4}$ and the recovery rate $\gamma=1$ (i.e.\,an infected node only spread the epidemic in one time step). For epidemics on the fully connected network, the local infection rate $\beta_1=6\times10^{-3}$. And for epidemics on the random and scale-free networks, $\beta_1=0.8$. \reffig{fig-threshold} shows that the  final outbreak size predicted by \refeq{eq-rinf0} is in close agreement with simulation results. The hybrid epidemics on the random network and the scale-free network exhibit similar outbreak sizes for large values of $\alpha$. It is also evident that the hybrid epidemic is characterised by a phase change, where the threshold is well predicted by \refeq{eq:threshold}. 

\section*{Hybrid Spreading In A Metapopulation}

We now extend the above theoretical results for a single-population to analyse hybrid spreading in a metapopulation which consists of a number of subpopulations. Local infection happens only between nodes in the same subpopulation whereas global infection occurs both within and between subpopulations.
\subsection*{Hybrid Spreading At The Population Level\label{strategies}}

We define a subpopulation as susceptible if it contains only susceptible nodes. A subpopulation is  infected if it has at least one infected node. A subpopulation is recovered if it has at least one recovered node and all other nodes are susceptible. Only global spreading enables infection between subpopulations, whereas spreading within a subpopulation can occur via both local and global spreading. 

The final outbreak size at the population level $R_{\infty}$, is defined as the proportion of subpopulations that are recovered when the epidemic stops spreading.
We define that a subpopulation A directly infects another subpopulation B if an infected node in A infects a susceptible node in B. We define the population reproduction number, $R_p$, as the average number of other subpopulations that an infected subpopulation directly infects before it recovers. Note that our definition of $R_p$ is similar to $R_*$ in \cite{Colizza_Vespignani_2007} but the definition of a metapopulation in \cite{Colizza_Vespignani_2007} is different. In the simulations and theoretical analysis, we approximate $R_p$ as the population reproduction number of the initially infected subpopulation $p_0$, i.e. the average number of other subpopulations that $p_0$ directly infects. This approximation becomes exact when the metapopulation has infinite number of subpopulations each with the same network structure.
A metapopulation includes many subpopulations. In order for an epidemic to spread in a metapopulation, an infected subpopulation should infect at least one other subpopulation before it recovers, i.e. the threshold condition of the hybrid epidemic at the population-level is $R_p>1$.
%

\begin{figure}\centering\small
\begin{overpic}[width=0.8\textwidth]{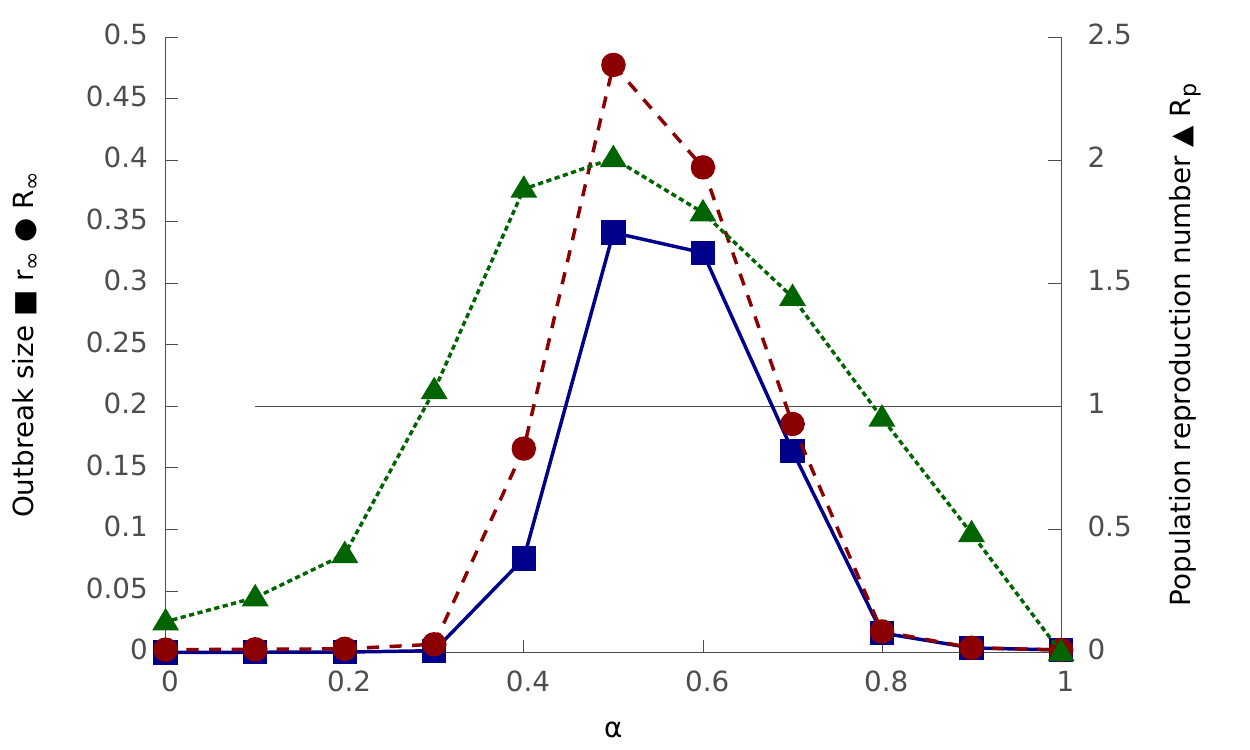}
 \put(0,55){(a)}
\end{overpic}
\begin{overpic}[width=0.8\textwidth]{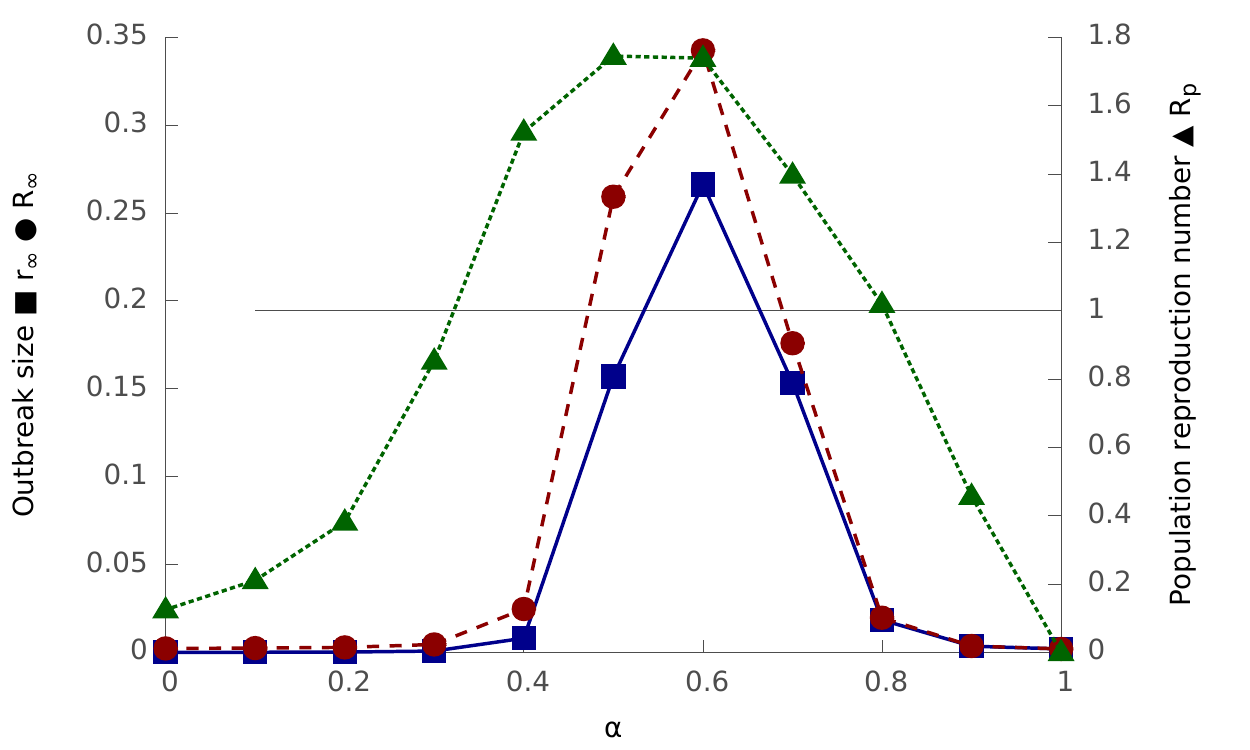}
 \put(0,55){(b)}
\end{overpic}
\caption{\label{fig-ob-RN0-a}Simulation results of hybrid epidemics in a  metapopulation where (a) each subpopulation is a random network and (b) each subpopulation is a scale-free network. Three quantities are shown as a function of the hybrid tradeoff $\alpha$, including the final outbreak size as the fraction of recovered nodes $r_{\infty}$ (squares);  the final outbreak size as the fraction of recovered subpopulations $R_{\infty}$ (circles); and the population reproduction number, $R_p$ (triangles, right y-axis).  The metapopulation contains 500 subpopulations each  with 100 nodes. In (a) each subpopulation is a random network with an average degree of 5; and In (b) each subpopulation is a scale-free network with a power-law degree distribution $p_k \sim 2m^2k^{-3}$ which is generated by the configuration model \cite{Newman_Book_2010} with the minimum degree $m=3$.. The  local infection rate $\beta_1=0.8$, the  global infection rate $\beta_2=10^{-6}$ and the recovery rate $\gamma=1$. Initially 3 random nodes in a subpopulation are infected. Simulation results are shown as points and each result is averaged over 1000 runs.}
\end{figure}

We conduct epidemic simulations on a  metapopulation containing 500 subpopulations each with 100 nodes. Two topologies for local spreading in each subpopulation are considered: random network and scale-free network. \reffig{fig-ob-RN0-a} shows simulation results of the final outbreak sizes $r_{\infty}$ and $R_{\infty}$ and the population reproduction number $R_p$ (right y axis)  as a function of the hybrid tradeoff $\alpha$. Epidemic parameter values are included in \reffig{fig-ob-RN0-a}'s legend. For both the random and scale-free networks, all three functions show a bell shape curve regarding $\alpha$. It is clear that the epidemic will not cause any significant infection if it uses only local spreading ($\alpha=1$) or only global spreading ($\alpha=0$). For the random network, the maximal outbreak at the node level $r^*_{\infty}=0.34$ is obtained around the optimal hybrid tradeoff $\alpha^*=0.5$. That is, if 50\% of the infection events occur via local spreading (and the rest via global spreading), the epidemic will ultimately infect 34\% of all nodes in the metapopulation. At the population level, the total percentage of recovered subpopulations $R_{\infty}$ follows a very similar trend to $r_{\infty}$, and the maximum epidemic size in terms of subpopulations occurs at the same optimal $\alpha^*$. The population reproduction number $R_p$ follows a similar trend to the final outbreak sizes $R_{\infty}$ and $r_{\infty}$. The threshold $R_p>1$  defines the range of $\alpha$ for which the final outbreak sizes are significantly larger than zero.

It is important to appreciate that although the maximal $R_p^*$ is uniquely defined by the optimal $\alpha^*$, other $R_p$ values can be obtained by {\it two} different $\alpha$ values, on either side of the optimal $\alpha^*$, potentially representing different epidemic dynamics. As the hybrid epidemic for random and scale-free networks exhibit similar properties, for simplicity we only show results for the random network in the following.

\subsection*{Prediction of the Population Reproduction Number $R_p$ }
The population reproduction number $R_p$ is a fundamental characteristic of hybrid epidemics in a metapopulation.
%
We consider a metapopulation with $N+1$ subpopulations, which are denoted as $p_i$ where $i=0,1,2...N$. Each subpopulation has $n$ nodes connected to a same structured local spreading network . $p_0$ is the subpopulation where the epidemic starts from.

We assume the infection inside the initially infected subpopulation $p_0$ is all caused by infected nodes inside $p_0$. That is, we neglect the effects of global spreading of other $N$ subpopulations on $p_0$. This is an acceptable assumption when the metapopulation has a larger number of subpopulations. Under these conditions, hybrid spreading within $p_0$ is the same as spreading in a single-population, which has been analysed in previous sections. To predict $R_p$, we first analyse the expected number of nodes outside $p_0$ that will be infected by $p_0$. We then estimate the number of other subpopulations that these infected nodes should belong to. Let $s_N(t)$ represent the probability that a random test node in other subpopulations are susceptible at time $t$. Using the same parameters defined in the analysis about hybrid epidemics in a single population, we have $s_N(t) = \vartheta(t)^n$ where $n$ is the number of node in $p_0$. 

\begin{figure}
\centering\small 
\begin{overpic}[width=0.8\textwidth]{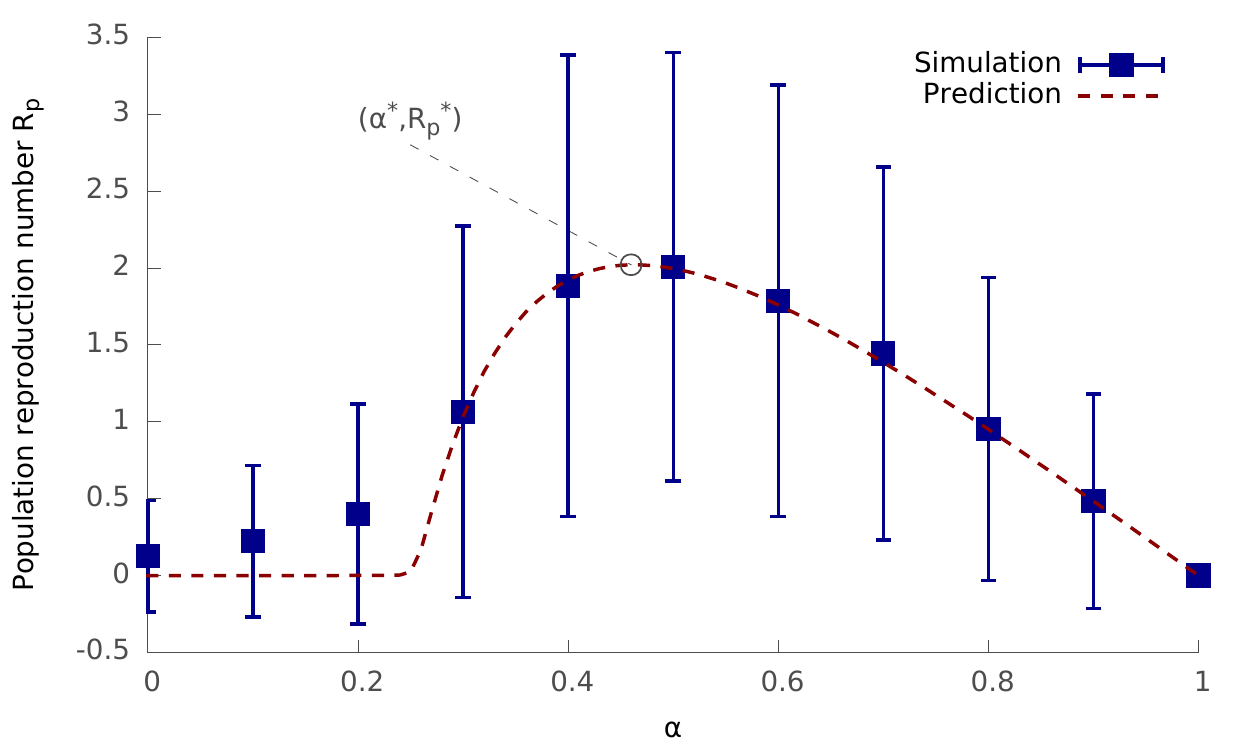}
\end{overpic}
\caption{\label{fig-a-RN0}The population reproduction number $R_p$ as a function of the hybrid tradeoff $\alpha$. Theoretical predictions from \refeq{eq:R_N0m} are shown as a dashed curve. Simulation results are shown as points (average over 1,000 runs) and bars (one standard deviation). The metapopulation and epidemic parameters are the same as \reffig{fig-ob-RN0-a}a.}
\end{figure}

When $p_0$ recovers at time $T$, the {\it fraction} of nodes in other subpopulations that have been infected by (infected nodes in) $p_0$ (via global spreading) is $x_N=1-s_N(T)=1-\vartheta(T)^n=1-w(\theta_T)^n$ where we have used \refeq{eq-vartheta}. Then the {\it number} of such infected nodes is
\begin{equation}
X_N=x_NnN=(1-w(\theta_T)^n)nN
\end{equation}
where $nN$ is the total number of nodes in other $N$ subpopulations and $\theta_T$ can be numerically calculated as $\theta_\infty$ by fixed-point iteration of \refeq{eq-theta}. As the nodes are infected randomly via the global spreading, the probability that an infected node does not belong to a particular subpopulation $i$ is $1-1/N$; and the probability that none of these infected nodes belongs to the subpopulation $i$ is $(1-1/N)^{X_N}$. So the probability that at least one infected node belongs to the subpopulation $i$ is $1-(1-1/N)^{X_N}$. Thus the population reproduction number $R_p$, which is the number of other subpopulations that these infected nodes should belong to, is:
\begin{equation}\label{eq:R_N0m}
R_p=N(1-(1-{1}/{N})^{X_N})
\end{equation}
 \reffig{fig-a-RN0} compares the predicted $R_p$ against simulation results as a function of the hybrid tradeoff $\alpha$. $R_p$ is characterised by a bell-shaped curve. It peaks at the optimal hybrid tradeoff $\alpha^*$ where the population reproduction number achieves its maximal value $R_p^*$. This optimal point is of particular interest as it represents the optimal trade-off between the two spreading mechanisms, where the hybrid epidemic is most infectious and therefore has the most extensive outbreak. 

\subsection*{The Optimal Hybrid Tradeoff $\alpha^*$ and the Maximal $R_p^*$ \label{sec:netscase}}
We next investigated the maximum epidemic outbreak in the context of varying infectivity and recovery rates. For a given set of epidemic variables, we calculate the theoretical prediction of $R_p$ as a function of $\alpha$ using \refeq{eq:R_N0m}, and then we obtain the optimal $\alpha^*$  and the maximal $R_p^*$. For ease of analysis, we fix the global infection rate $\beta_2$ at a small value of $10^{-6}$ and then focus on the local infection rate $\beta_1$ and the recovery rate $\gamma$.

\begin{figure}\centering\small
\begin{overpic}[width=0.49\textwidth]{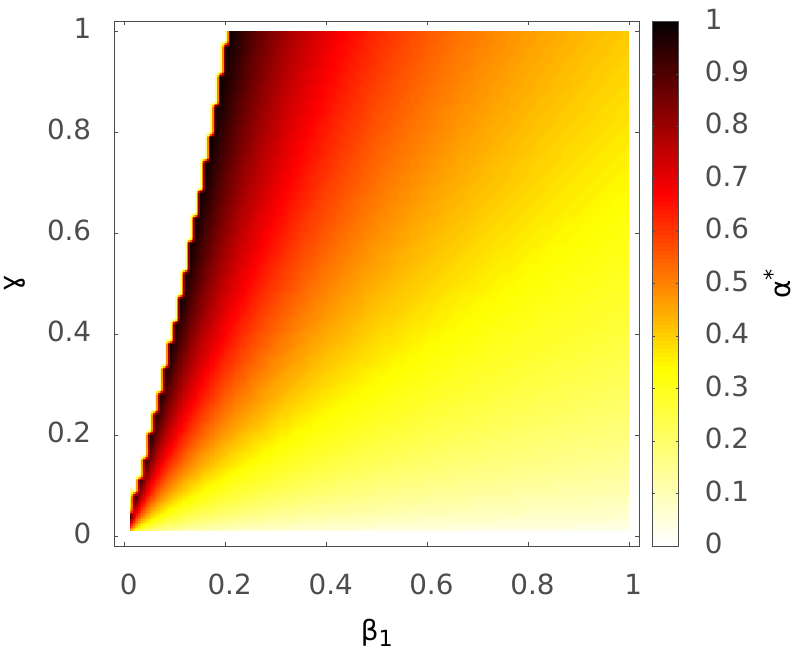}
 \put(0,75){\textbf{(a)}}
\end{overpic}
\begin{overpic}[width=0.49\textwidth]{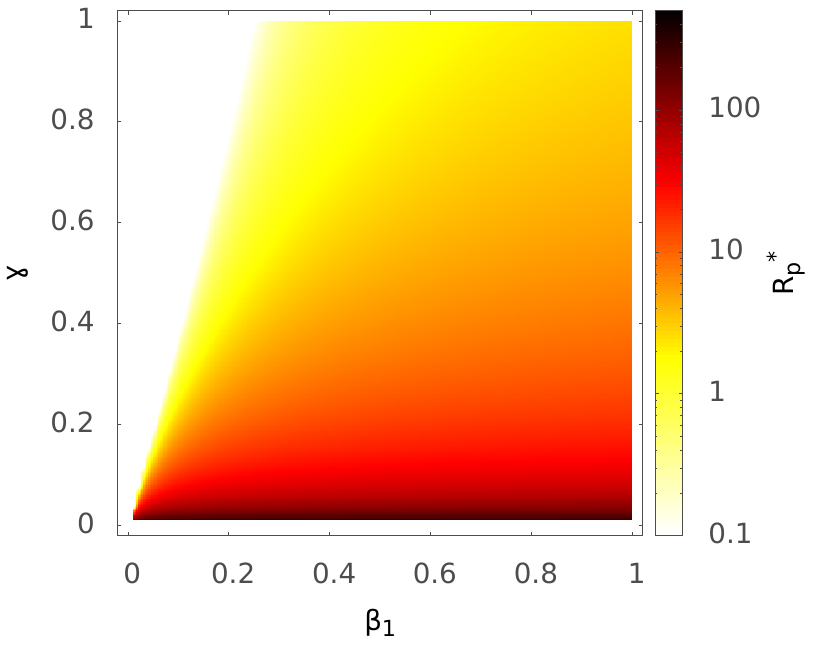}
 \put(0,75){\textbf{(b)}}
\end{overpic}\\
\begin{overpic}[width=0.8\textwidth]{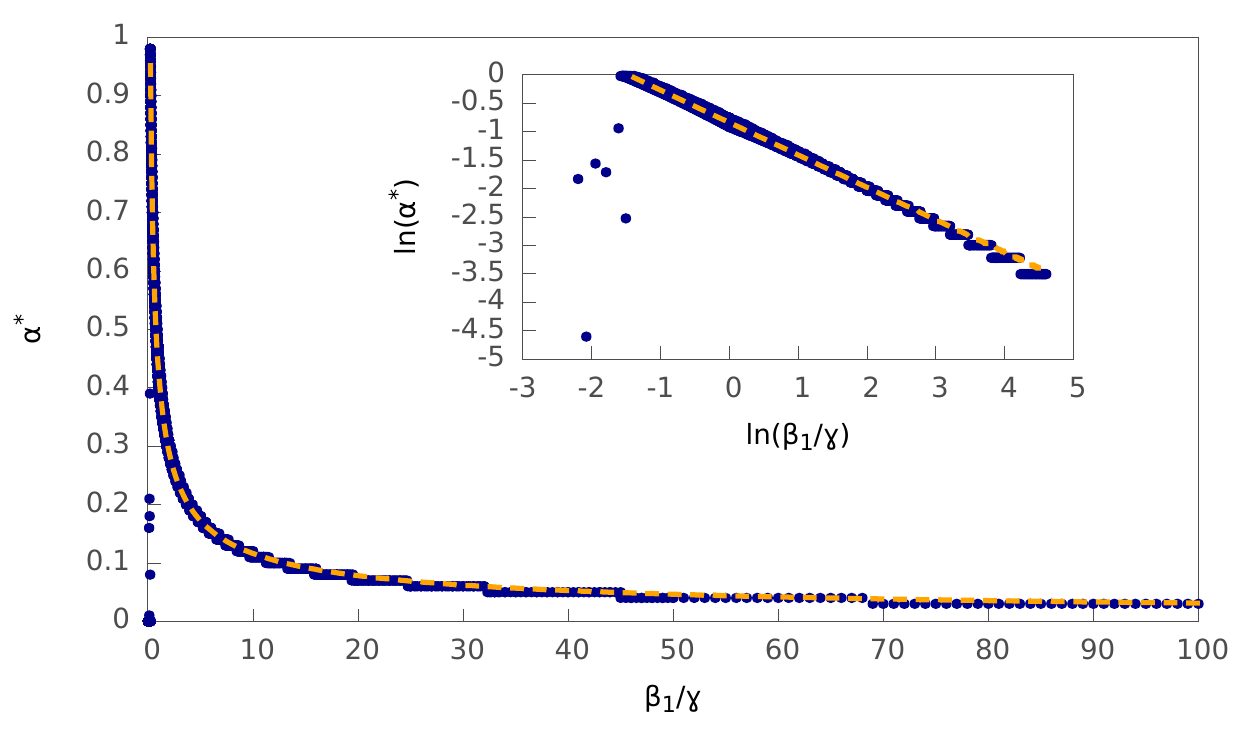}
 \put(0,55){\textbf{(c)}}
\end{overpic}
\begin{overpic}[width=0.49\textwidth]{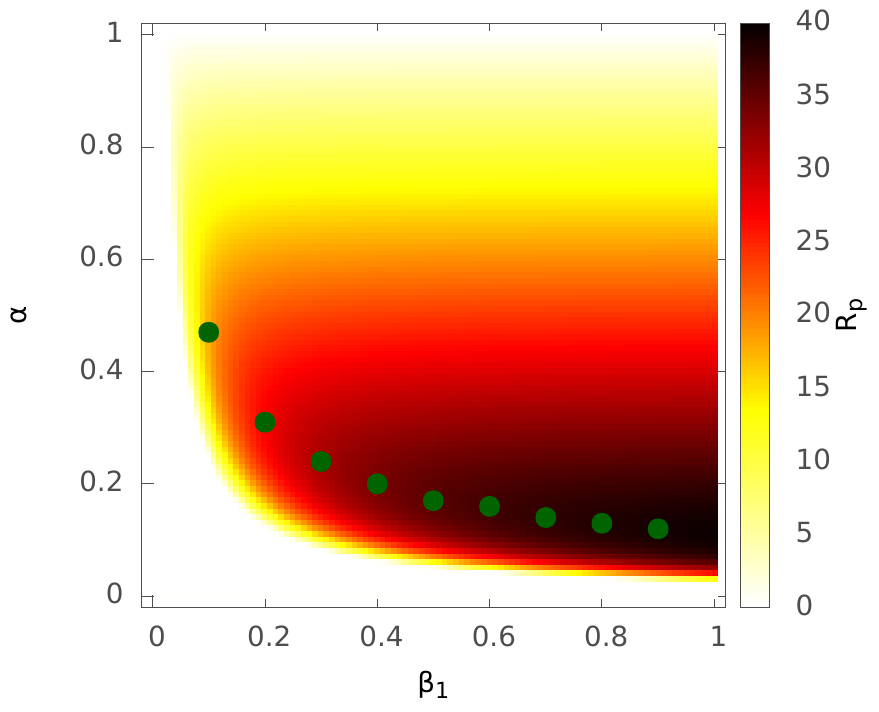}
 \put(0,75){\textbf{(d)}}
\end{overpic}
\caption{\label{fig-mamRc}The estimated optimal hybrid tradeoff $\alpha^*$ and the maximal population reproduction number $R_p^*$ for hybrid epidemics in a  metapopulation. (\textit{a}) $\alpha^*$ as a function of  local infection rate $\beta_1$ and recovery rate $\gamma$; (\textit{b})~ $R_p^*$ as a function of $\beta_1$ and $\gamma$; (\textit{c}) $\alpha^*$ as a function of $\beta_1/\gamma$, which is fitted by a dash line of $ln(\alpha^*)=-0.84-0.57\cdot ln(\beta_1/\gamma)$. (\textit{d}) Population reproduction number $R_p$ as a function of $\alpha$ and $\beta_1$  with $\gamma=0.1$, where the points are the corresponding optimal $\alpha^*$ for given $\beta_1$. We fix $\beta_2=10^{-6}$ and the metapopulation is as in \reffig{fig-ob-RN0-a}a.}
\end{figure}

\reffig{fig-mamRc}a shows the optimal hybrid tradeoff $\alpha^*$ as a function of $\beta_1$ and $\gamma$. For a given $\gamma$, a larger $\beta_1$ results in a smaller $\alpha^*$. Intuitively this can be understood as  when the efficiency of local spread increases, less effort needs to be devoted to this spreading mechanism, and more can be allocated to global spreading. 
On the other hand, for a given $\beta_1$, a larger $\gamma$ results in an increase in $\alpha^*$. When the recovery rate is higher, nodes remain infectious for shorter times. In this case, in order to achieve the maximum epidemic outbreak, more local infection is favoured, since this will allow an infected subpopulation to remain infected for longer, and hence increase the probability of infecting other subpopulations before it recovers.
A plot of $\alpha^*$ versus $\beta_1/\gamma$ is shown in \reffig{fig-mamRc}c. The fitting on a log-log scale in the inset indicates the two quantities have a power-law relationship, i.e. $\alpha^*$ is determined by $\beta_1/\gamma$. This means the optimal hybrid tradeoff $\alpha^*$ can be predicted when $\beta_1/\gamma$ is known.

\reffig{fig-mamRc}b shows the maximal $R_p^*$ as a function of $\beta_1$ and $\gamma$, where the $R_p^*$ is obtained when the corresponding value of $\alpha^*$ in \reffig{fig-mamRc}a is used.  $R_p^*$ is very sensitive to the recovery rate $\gamma$. As $\gamma$ approaches zero, the value of $R_p^*$ increases dramatically (note that $R_p^*$ uses a log-scale colour-map) regardless of value of $\beta_1$. This is in agreement with the intuition that a low recovery rate will favour any type of epidemic spreading. For a fixed $\gamma$, $R_p^*$ increases with $\beta_1$. An increased infection rate of local spreading will obviously increase the reproductive number, if other parameters are kept constant, but the effect is much smaller than that of changing the recovery rate, because global spreading maintains the reproductive number when local spreading falls to low values. 

\reffig{fig-mamRc}a shows a clear phase shift between areas where an epidemic occurs (the coloured area) and areas where it does not (the white area towards the top-left corner). Accordingly, the corresponding $R_p^*$ in \reffig{fig-mamRc}b in the area where no epidemic occurs is very small. The boundary between the epidemic and non-epidemic phase space is defined by the line $\beta_1/(\beta_1+\gamma-\gamma\beta_1)\approx0.2$. This is the threshold for completely local spreading in a single-population: $\beta_1/(\beta_1+\gamma-\gamma\beta_1)>g_0'(1)/g_0''(1)$ and $g_0'(1)/g_0''(1)\approx0.2$ for the network topology used. Since the global infection rate $\beta_2$ is fixed at a small value, no major spreading will occur either within or between subpopulations below this threshold.

\reffig{fig-mamRc}d plots $R_p$ as a function of $\beta_1$ and $\alpha$ on a log-log scale while fixing $\gamma=0.1$. For given values of $\beta_1$, the corresponding optimal $\alpha^*$ are shown as points. 
We can see that points always fall in the area of the maximal $R_p^*$ for the given $\beta_1$. Each point represents a local optimum. The global optimum, the largest possible value of $R_p$, is obtained towards the bottom-right corner, where the local infection rate is high but the epidemic spends most effort on global spreading. Infection across subpopulations can only be achieved by global spreading. Since global spreading has a low infection rate, the epidemic should spend most of its time (or resource) on global spreading. There will be much less time spent on local spreading but its infection rate is high anyway.

\subsection*{}

%
%

\section*{Discussion} 

Hybrid spreading, the propagation of infectious agents using two or more alternative mechanisms, is a common feature of many real world epidemics. Widespread epidemics (e.g. computer worms) typically spread efficiently by local spreading through connections within a subpopulation, but also use global spreading to probe distant targets usually with much lower infectivity. In many cases, the amount of resources (e.g. time, energy or money) which an infectious agent can devote to each mode of propagation is  limited. This study  focuses on the tradeoff between local and global spreading, and the effect of this tradeoff on the outbreak of an epidemic.

We develop a theoretical framework for investigating the  relationships between $\alpha$, the relative weight given to each spreading mechanisms, and the other epidemic properties. These properties include epidemic infectivity, subpopulation structure, epidemic threshold, and population reproduction number. The predictions of the theoretical model agree well with stochastic simulation results, both in single populations and in metapopulations.

Our analysis shows that epidemics spreading in a metapopulation may be critically hybrid epidemics where a combination of the two spreading mechanisms is essential for an outbreak and neither completely local spreading nor completely global spreading can allow epidemics to propagate successfully. 

Our study reveals that, in metapopulations, there exists an optimal tradeoff between global and local spreading, and provides a way to calculate this optimum given information on other epidemic parameters. These results are supported by our recent study \cite{Zhang_2014} on measurement data of the Internet worm Conficker \cite{Shin_2012, CAIDA_2008_aft, CAIDA_2008_bfr}.

The above results are of practical relevance when the total amount of time or capacity that is allocated to spreading is  limited by some resource constraint. For example, the total probing frequency of computer worms is often capped at a low rate to prevent them from being detected by anti-virus software.  Furthermore, other epidemic parameters, such as  local or global infection rates are difficult to change because they derive from inherent properties of the infectious agent. For example it would be difficult to increase the  global infection rate of an Internet worm. The tradeoff between different types of spreading therefore becomes a key parameter in terms of design strategy, which can be manipulated to maximise outbreak size. 

The consideration of hybrid spreading mechanisms also has some interesting implications for strategies for protecting against the spread of epidemics. It is clear from both theoretical considerations and simulations  that epidemics can spread with extremely low global infection rates (far below individual recovery rates), provided there is efficient local infection. Such conditions are common for both cyber epidemics (as computers within infected local networks tend to be more vulnerable to infection \cite{Zou_2006}) and in infectious disease epidemics, where contacts between family or community members are often much closer and more frequent than the overall population. Protection strategies which target local networks collectively (for example intensive local vaccination around individual disease incidents, as was used in the final stages of smallpox eradication \cite{DeQuadros_1972}) may therefore be a key element of future strategies to control future  mixed spreading epidemics.

In conclusion, our study highlights the importance of the tradeoff between local and global spreading, and manipulation of this tradeoff may provide a way to improve strategies for spreading, but also a way to estimate the worst outcome (i.e. largest outbreak) of hybrid epidemics which can pose serious threats to Internet security .

\section*{Methods}

\subsection*{Threshold for local spreading using Newman's method}

Here we use Newman's method \cite{Newman_2002} to obtain the threshold condition for the local spreading. Firstly we need to calculate the ``transmissibility'' $T$ which is the average probability that an epidemic is transmitted between two connected nodes, of which one is infected and the other is susceptible. According to \cite{Newman_2002}, for the discrete time case $T$ can be calculated as 
\begin{equation}\label{eq-T0}
T=1-\int_0^\infty d\beta_1 \sum_{\tau=0}^\infty p(\beta_1) p(\tau) (1-\beta_1)^\tau
\end{equation}
where $\tau$ is the time steps that an infected node remains infected, $p(\tau)$ and $p(\beta_1)$ respectively are the probability distribution of $\tau$ and $\beta_1$. For the model in this paper, $\beta_1$ is a constant and $p(\tau)=(1-\gamma)^{\tau-1}\gamma$, in which $(1-\gamma)^{\tau-1}$ is the probability that an infected node has not recovered until $\tau-1$ steps after infection, and $\gamma$ is the probability that the node recovers at the $\tau$th step after infection. Also for the model in this paper, each infected node at least remains infected for $1$ time step. So that $T$ for our model can be obtained as
\begin{equation}\label{eq-T1}
T=1-\sum_{\tau=1}^\infty (1-\gamma)^{\tau-1}\gamma (1-\beta_1)^\tau=\frac{\beta_1}{\beta_1+\gamma-\gamma\beta_1}
\end{equation}
According to \cite{Newman_2002} the epidemic threshold for completely local spreading is $Tg_0''(1)/g_0'(1)>1$ i.e. $\beta_1g_0''(1)/[g_0'(1)(\beta_1+\gamma-\gamma\beta_1)]>1$. This is the same as the epidemic threshold for completely local spreading obtained in this paper.

Note that treating each edge as having this value of $T$ independently will lead to the correct epidemic threshold and final size calculation, but there are further discussions on its correctness in calculating the infection probabilities \cite{Miller_2007, Hastings_2006,Kenah_2007,Miller_2008}.

\subsection*{Simulation settings}

Random networks used in all simulations have a Poisson degree distribution and they are generated by the Erd\H{o}s-R\'{e}nyi (ER) model \cite{ER_1959} with the average degree of 5.

Scale-free networks used in all simulations have a power-law degree distribution $p_k \sim 2m^2k^{-3}$ and they are generated by the configuration model \cite{Newman_Book_2010} with the minimum degree $m=3$.

\reffig{fig-threshold} - simulations in a single-population: $\bullet$ Size of single-population: 1,000 nodes; $\bullet$ Single-population topology: fully connected network, random network and scale-free network; $\bullet$ Local infection rate: $\beta_1=0.8$ (except for fully connected network $\beta_1=6\times10^{-3}$); $\bullet$ Global infection rate: $\beta_2=10^{-4}$; $\bullet$ Recovery rate: $\gamma=1$; $\bullet$ Initial condition: all nodes are susceptible except 5 randomly-chosen nodes are infected; $\bullet$ Number of simulation runs averaged for each data point: 1,000.

\reffig{fig-ob-RN0-a} - simulations in a metapopulation: $\bullet$ Size of metapopulation: 500 subpopulations each with 100 nodes; $\bullet$ Subpopulatin topology: random networks and scale-free networks;  $\bullet$ Local infection rate: $\beta_1=0.8$; $\bullet$ Global infection rate: $\beta_2=10^{-6}$; $\bullet$ Recovery rate: $\gamma=1$;  $\bullet$ Initial condition: all nodes are susceptible except 3 randomly-chosen nodes are infected; $\bullet$ Number of simulation runs averaged for each data point: 1,000.

\reffig{fig-a-RN0} and \ref{fig-mamRc} - theoretical predictions about hybrid epidemics: Same as \reffig{fig-ob-RN0-a} except only the random network topology is considered.


\section*{Acknowledgments}
We thank Prof. Valerie Isham of UCL for her helpful comments. C.Z. was supported by the Engineering and Physical Sciences Research Council of UK (No. EP/G037264/1), the China Scholarship Council (File No. 2010611089), and the National Natural Science Foundation of China (Project no. 60970034, 61170287, 61232016). I.J.C. acknowledges support from the EPSRC IRC in Early Warning Sensing Systems for Infectious Diseases (grant reference EP/K031953/1). J.C.M. was supported in part by the RAPIDD program of the Science and Technology Directorate, Department of Homeland Security and the Fogarty International Center, National Institutes of Health. The content is solely the responsibility of the authors and does not necessarily represent the official views of the National Institute of General Medical Sciences or the National Institutes of Health. The funders had no role in study design, data collection and analysis, decision to publish, or preparation of the manuscript.

\section*{Author contributions}

C.Z., S.Z., I.J.C., and B.M.C. designed the study. C.Z and J.C.M. conducted the mathematical modelling and derivation. C.Z. performed the computational analysis and simulations. S.Z. and B.M.C. wrote the manuscript with contributions from C.Z. and J.C.M. and I.J.C.

\section*{Additional information}

\textbf{Competing financial interests:} The authors declare no competing financial interests.

\end{document}